# Unconventional critical behaviour in a quasi-two-dimensional organic conductor

F. Kagawa[1], K. Miyagawa[1,2] & K. Kanoda[1,2]

[1]*Department of Applied Physics, University of Tokyo, Bunkyo-ku, Tokyo 113-8656, Japan.* [2]*CREST, Japan Science and Technology Corporation, Kawaguchi 332-0012, Japan.*

**Changing the interactions between particles in an ensemble—by varying the temperature or pressure, for example—can lead to phase transitions whose critical behaviour depends on the collective nature of the many-body system. Despite the diversity of ingredients, which include atoms, molecules, electrons and their spins, the collective behaviour can be grouped into several families (called 'universality classes') represented by canonical spin models[1]. One kind of transition, the Mott transition[2], occurs when the repulsive Coulomb interaction between electrons is increased, causing wave-like electrons to behave as particles. In two dimensions, the attractive behaviour responsible for the superconductivity in high-transition temperature copper oxide[3,4] and organic[5–7] compounds appears near the Mott transition, but the universality class to which two-dimensional, repulsive electronic systems belongs remains unknown. Here we present an observation of the critical phenomena at the pressure-induced Mott transition in a quasi-two-dimensional organic conductor using conductance measurements as a probe. We find that the Mott transition in two dimensions is not consistent with known universality classes, as the observed collective behaviour has previously not been seen. This peculiarity must be involved in any emergent behaviour near the Mott transition in two dimensions.**

Using the pressure-sweep technique to vary the effective electronic interactions in $(V_{1-x}Cr_x)_2O_3$, Limelette *et al.* successfully revealed the critical phenomena of the Mott transition in three dimensions (3D; ref. 8). The critical exponents are consistent with those found using mean-field theory. For the study of the Mott transition in two dimensions (2D), organic conductors, which are low-dimensional and highly compressible, are suitable. The organic family with a half-filled band[9], $\kappa$-(BEDT-TTF)$_2$X, are good candidates, where BEDT-TTF is bis(ethylenedithio)tetrathiafulvalene and X stands for various kinds of anions. The $\kappa$-(BEDT-TTF)$_2$X family has a quasi-2D layered structure (Fig. 1a) and shows similar behaviour to high-transition temperature (high-$T_c$) copper oxides[10]. In the conceptual phase diagram of the family[11] (Fig. 1b), unconventional superconductivity appears in the marginal metallic phase near the antiferromagnetic insulator, like high-$T_c$ copper oxides[3,4], although the Mott transition is induced by pressure in the organics but by carrier doping in the copper oxides. Furthermore, the exotic pseudogap behaviour is emergent near the Mott transition in both the organics[12] and copper oxides[3,4]. These similarities suggest that they share common physics—the Mott transition in 2D correlated electrons.

The Mott insulator κ-(BEDT-TTF)$_2$Cu[N(CN)$_2$]Cl (denoted by κ-Cl hereafter) undergoes an insulator to metal transition when pressure is applied and thus allows us to approach the critical endpoint of the first-order Mott transition[13–16] (Fig. 1b). This feature provides the foundation of our research because the critical phenomena, which are direct clues to identifying the universality class, are generally found around the endpoint, in analogy to the liquid–gas transition. In the Mott transition, the conductance in the metallic region was found to describe the critical phenomena very well by Limelette *et al.*[8]. To characterize the Mott critical phenomena of κ-Cl in detail, we measured the in-plane conductance $G$ around the endpoint with the standard d.c. four-probe method under isothermal pressure sweep (~3 MPa h$^{-1}$ in ascending and descending processes) using helium gas as a pressure medium. Throughout the experiment, the conductance was independent of the applied current within two orders of magnitude. The experiments were performed on three single crystals of κ-Cl, which were grown by the conventional electrochemical oxidation of BEDT-TTF[17].

Figure 1c shows the conductance profile as a function of temperature $T$ and pressure $P$. Below the critical temperature $T_c$, the conductance jump (shaded area) is seen at well-defined pressures, where the first-order metal–insulator transition occurs. Above $T_c$, however, the pressure dependence of conductance $G_T(P)$ at a given temperature is continuous, corresponding to the metal–insulator crossover. The characteristic 'crossover pressure' $P_{cross}(T)$ is defined as that giving the peak in the pressure derivative of conductance, $G'_T(P) \equiv \partial G_T(P)/\partial P$ (Fig. 1d). The first-order transition line and the crossover line are represented in the pressure–temperature phase diagram (Fig. 2a).

Around the critical endpoint, we found three regions of critical behaviour in the conductance, which we call regions (1)–(3). Two of them are visible in Fig. 1c: in region (1) at $T = T_c$, the $G_{T=T_c}(P)$ curve is continuous but has a vertical gradient at the critical pressure $P_c$ (see the red curve in Fig. 1c); and in region (2) below $T_c$, the conductance jump continuously vanishes as $T \to T_c$ from below (see the green curve in Fig. 1c). Region (3) of critical behaviour is found in Fig. 1d, where the peak value in the $G'_T(P)$ curve grows continuously and diverges as $T \to T_c$ from above (see the red curve in Fig. 1d). The fixed parameters of the critical endpoint are $T_c \approx 39.7 \pm 0.1$ K and $P_c \approx 24.8 \pm 0.1$ MPa, which are sample-independent.

Generally, the critical phenomena show power-law behaviour and are characterized quantitatively by the exponents of the power laws (called 'critical exponents'). As phase transitions belonging to the same universality class are characterized by the same set of critical exponents, we can identify the universality class from the values of the exponents. Below we present the results of the κ-Cl crystal investigated in the greatest detail. As seen in Fig. 2b–d (see also Methods), the conductance behaviour of regions (1)–(3) follows power laws around the critical endpoint. From the slopes in the logarithmic plots, we determined that the critical exponents ($\delta$, $\beta$, $\gamma$) are nearly equal to (2, 1, 1). In the data analysis, we avoided the region very close to the endpoint, $|T - T_c|/T_c < \sim 2\times10^{-2}$ and $P - P_c < \sim 0.2$ MPa, because the analysis in this region is easily affected by ambiguities in the determination of $T_c$ and $G_c \equiv G_{T=T_c}(P_c)$, the pressure accuracy ($\pm 0.05$ MPa) and the sample quality. The reproducibility of the power-law behaviour was ensured for two other crystals ($\sim 1.9 < \delta < \sim 2$, $\sim 0.9 < \beta < \sim 1$ and $\sim 0.9 < \gamma < \sim 1$) over the wide pressure–temperature range except in the close vicinity of the endpoint, although the prefactor of each power law (normalized by the value of $G_c$) showed some sample dependence ($\pm 30\%$).

It is surprising that the observed exponents, ($\delta$, $\beta$, $\gamma$) $\approx$ (2, 1, 1), are not only previously unknown but also far from any existing values (Fig. 3). The widely known universality classes have in common $\delta \geq 3$ and $\beta \leq 0.5$ (note that ($\delta$, $\beta$) = (3, 0.5) are the mean-field values). This is a natural consequence of collective (non-local) fluctuation effects, which generally cause more rapid ordering than the mean-field scheme dealing with the single-site (local) problem. In the present case, however, the deviation of $\delta$ and $\beta$ from the mean-field values is in the opposite direction (that is $\delta < 3$ and $\beta > 0.5$) to the conventional cases, that is, the ordering is more moderate than the mean-field scheme. This feature is obviously anomalous, suggesting the discovery of novel critical behaviour out of the conventional framework.

Below we examine whether the present values correctly correspond to critical behaviour, using two methods based on a scaling hypothesis[1]. One is the scaling relation $\delta = 1+(\gamma/\beta)$, which should be fulfilled by any critical phenomenon. It is evident that the present values, ($\delta$, $\beta$, $\gamma$) $\approx$ (2, 1, 1), satisfy the relation. The other prerequisite is the existence of a universal form of the equation of state[1,18], which in the present case is expected to follow (see Methods):

$$\{G_T(P) - G_T(P_{cross}(T))\}/\{P - P_{cross}(T)\}^{1/\delta} = f_+\left(\frac{P - P_{cross}(T)}{|T - T_C|^{\delta\beta}}\right) \quad \text{for } T > T_C, \quad (1)$$

$$\{G_T(P) - G_C\}/\{P - P_I(T)\}^{1/\delta} = f_-\left(\frac{P - P_I(T)}{|T - T_C|^{\delta\beta}}\right) \quad \text{for } T < T_C, \quad (2)$$

where $P_I(T)$ represents the pressure of the conductance jump in the ascending-pressure process. Using the present value, ($\delta$, $\beta$) $\approx$ (2, 1), and the whole data set of $G_T(P)$ in the metallic region ($P > P_I(T)$ for $T < T_c$ and $P > P_{cross}(T)$ for $T > T_c$), we plotted $\{G_T(P) - G_T(P_{cross}(T))\}/\{P - P_{cross}(T)\}^{1/\delta}$ versus $\{P - P_{cross}(T)\}/|T - T_c|^{\delta\beta}$ for $T > T_c$ and $\{G_T(P) - G_c\}/\{P - P_I(T)\}^{1/\delta}$ versus $\{P - P_I(T)\}/|T - T_c|^{\delta\beta}$ for $T < T_c$. The resulting plot reasonably leads to the two scaling curves over a wide range (Fig. 4). The consistency between Figs 2 and 4 justifies the present exponents and the above analysis of the data (see also Supplementary Discussion). It is also confirmed that the scaling plot with ($\delta$, $\beta$) $\approx$ (2, 1) is well reproduced for two other crystals. Therefore we conclude that the observed exponents, $(\delta, \beta, \gamma) \approx (2, 1, 1)$, characterize correctly the critical behaviour of the (pressure-induced) Mott transition in a quasi-2D system; they are inconsistent with known universality classes and suggest novel critical behaviour. As seen above, the conductance describes the Mott criticality in κ-Cl well, as in $(V_{1-x}Cr_x)_2O_3$ (ref. 8), in that it shows the critical behaviour and satisfies the scaling hypothesis. This suggests that conductance is an essential quantity characterizing the Mott criticality.

The present results mean that 2D correlated electrons form a many-body system with anomalous collective behaviour, which cannot be understood from known spin models. Generally, the universality class depends on the system dimensionality, order-parameter dimensionality, and interaction range. We note that the Mott transition in the 3D system $(V_{1-x}Cr_x)_2O_3$ was reported[8] to be equivalent to the predicted conventional Ising transition in 3D[19,20]. Thus the two-dimensionality of κ-Cl apparently seems to be responsible for the unconventionality ($\delta < 3$ and $\beta > 0.5$). However, at least within the Ising scheme, two-dimensionality would give merely the conventional 2D-Ising (or mean-field) values. Therefore some new framework beyond the conventional ones should be invoked to explain the unconventional exponents, and it will be a challenging future problem. We note that recent theoretical work highlighting the quantum nature of electrons predicted Mott criticality with unconventional exponents in 2D (refs 21, 22).

Our conductance results suggest that novel criticality—perhaps belonging to a new universality class—is inherent in quasi-2D, giving rise to exotic properties such as unconventional superconductivity and pseudogap behaviour. This raises a fundamental problem of how this peculiarity of the system background is involved in the emergence of the macroscopic condensate near the Mott transition.

**Methods**

**Order parameter and scaling variables of the Mott transition**

Following the scaling hypothesis[1], we introduce three variables for the analysis of the critical phenomena: order parameter, $\phi$, and two scaling variables, $h$ and $\varepsilon$, where $h$ is the conjugate field of $\phi$ and $\varepsilon$ represents the distance to the critical endpoint. In the scaling framework, the critical exponents $(\delta, \beta, \gamma)$ around the endpoint, $(h, \varepsilon)=(0, 0)$ in the $h$–$\varepsilon$ plane, are defined as $\phi(h, 0) \propto h^{1/\delta}$, $\phi(0, \varepsilon) \propto |\varepsilon|^\beta$ and $\partial\phi(h,\varepsilon)/\partial h|_{h=0} \propto |\varepsilon|^{-\gamma}$. Note that these critical exponents are not independent of each other, but the scaling relation $\delta = 1 + (\gamma/\beta)$ should be fulfilled among them. To analyse the Mott critical phenomena, one should relate $(\phi, h, \varepsilon)$ of the Mott transition to the observable quantities. According to the theoretical framework of strongly correlated electrons, dynamical mean-field theory (DMFT)[18,23,24], the order parameter of the Mott transition has a scalar nature[18,20,24] and can be measured by the conductance in the metallic region, $G$–$G_c$ (refs 18, 24, 25). This has been demonstrated by an experiment on a 3D oxide[8]. In the

(pressure-induced) Mott transition, the experimental variables are pressure and temperature. However, the scaling valuables $h$ and $\varepsilon$ do not generally have direct correspondence to pressure and temperature, respectively, but are approximated well by a linear combination of $(T-T_c)$ and $(P-P_c)$ near the endpoint (linear-mixing approximation, see Supplementary Discussion): $h \approx (P-P_c)$ with some admixture of $(T-T_c)$ and $\varepsilon \approx (T-T_c)$ with some admixture of $(P-P_c)$. Following the approach adopted in the case of the liquid–gas transitions[1] or the Mott transition in a 3D system[8], we assume tentatively that the admixtures are negligibly small (non-mixing approximation). Thus we evaluated the critical exponents ($\delta$, $\beta$, $\gamma$) by taking ($G-G_c$, $P-P_c$, $T-T_c$) for ($\phi$, $h$, $\varepsilon$), that is, the exponents are determined from the power-law behaviour of conductance (1)–(3) (see main text) as follows: (1) $G_{T=T_c}(P)-G_c \propto (P-P_c)^{1/\delta}$, (2) $G_{met}(T)-G_c \propto (T_c-T)^{\beta}$ (along the first-order transition line), and (3) $\partial G_T(P)/\partial P|_{P=P_{cross}(T)} \propto (T-T_c)^{-\gamma}$ (along the crossover line), where $G_{met}(T)$ is the conductance value of the metallic regime just after the conductance jump in the ascending-pressure process. As seen in the text, the Mott criticality is well described by using this non-mixing approximation. In the Supplementary Discussion, we prove that the non-mixing approximation does not influence the present evaluation of the exponents.

**Universal form of equation of state**

Around the critical endpoint, the order parameter $\phi$ should follow a universal form of the equation of state including the critical exponents: $\phi(h,\varepsilon)/h^{1/\delta} = f_{\pm}(h/|\varepsilon|^{\delta\beta})$, where $f_+(x)$ for $\varepsilon > 0$ and $f_-(x)$ for $\varepsilon < 0$ are called 'scaling functions' depending on the details of the system. Within the non-mixing approximation, $\varepsilon \approx T-T_c$. Since $h = 0$ on the crossover line and the first-order transition line, $h$ for $T \neq T_c$ is given as follows: $h \approx (P-P_{cross}(T))$ for $T > T_c$ and $h \approx (P-P_1(T))$ for $T < T_c$, where $P_1(T)$ represents the pressure of the conductance jump in the ascending-pressure process. Above $T_c$, $\phi$ should be zero at $h = 0$. Thus $\phi$ for $T > T_c$ is given as $\phi \approx (G_T(P)-G_T(P_{cross}(T)))$. As a result, the universal form of equation of state is approximated as equations (1) and (2) (see main text). In fact, the successful scaling behaviour shown in Fig. 4 assures us that the non-mixing approximation is valid in the present case.

**Supplementary Information** is linked to the online version of the paper at www.nature.com/nature.

**Acknowledgements** We thank M. Imada, T. Itou, S. Miyashita, N. Nagaosa, S. Onoda, Y. Shimizu and N. Todoroki for discussions.

**Author Information** Reprints and permissions information is available at npg.nature.com/reprintsandpermissions. The authors declare no competing financial interests. Correspondence and requests for materials should be addressed to K.K. (kanoda@ap.t.u-tokyo.ac.jp).

8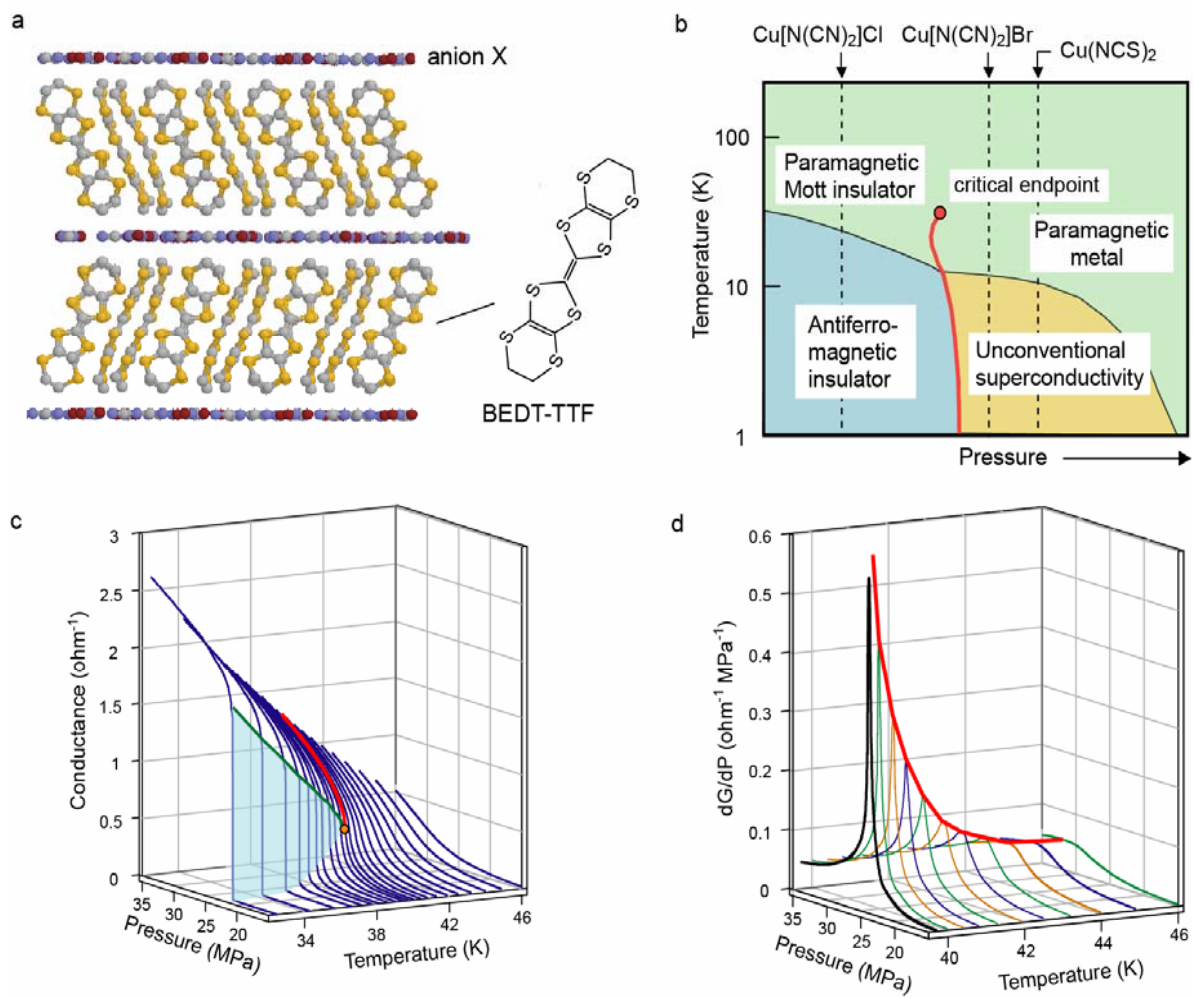

**Figure 1 | Crystal structure and conductance profile around the critical endpoint of a Mott transition. a**, Crystal structure of κ-(BEDT-TTF)$_2$Cu[N(CN)$_2$]Cl. 2D conducting layers of BEDT-TTF dimers are separated by insulating anion layers, like high-$T_c$ copper oxides. The dimer has one hole, and thus the band is half-filled. **b**, Generic pressure–temperature phase diagram of κ-(BEDT-TTF)$_2$X, of which three members (X=Cu[N(CN)$_2$]Cl, Cu[N(CN)$_2$]Br and Cu(NCS)$_2$) are indicated along the pressure axis. The red line represents the first-order transition, which terminates at a finite-temperature critical endpoint (filled circle). **c**, Pressure dependence of conductance $G_T(P)$ around the critical endpoint (filled circle). The shaded area indicates the conductance jump. The red and green curves represent the critical behaviour at $T = T_c \approx 39.7$ K and $T < T_c$, which give the critical exponents $\delta$ and $\beta$, respectively (see Fig. 2b and c). The hysteresis of the conductance jump (for example, ~0.2 MPa at ~32 K) is not appreciable at this scale. **d**, Pressure derivative of conductance $G'_T(P) \equiv \partial G_T(P)/\partial P$ as a function of pressure at temperatures above $T_c \approx 39.7$ K. The black curve shows the data at $T = T_c$. The red curve represents the critical divergence of the pressure derivative, which gives the critical exponent $\gamma$ (see Fig. 2d).

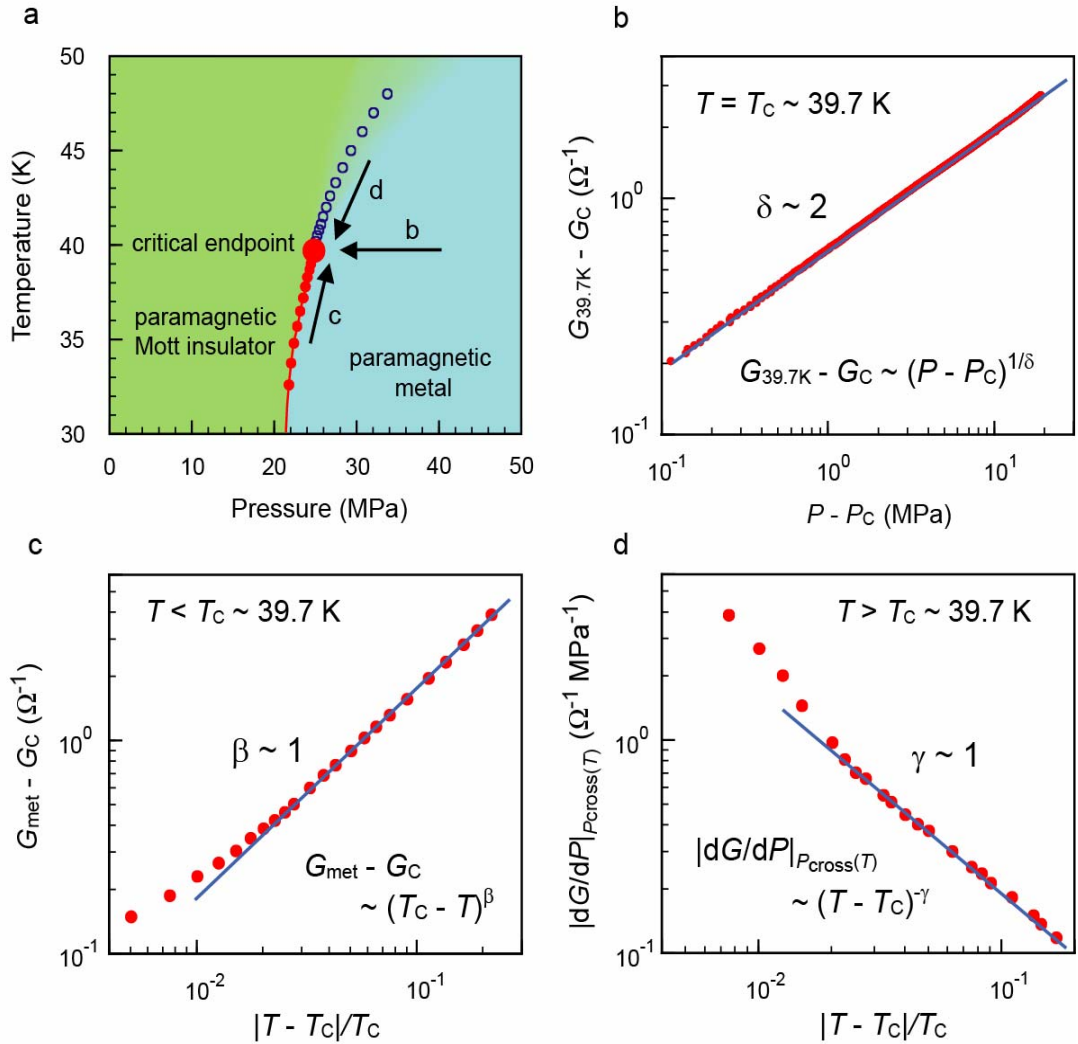

**Figure 2 | Critical exponents of the Mott transition in κ-(BEDT-TTF)$_2$Cu[N(CN)$_2$]Cl. a**, Pressure–temperature phase diagram of κ-(BEDT-TTF)$_2$Cu[N(CN)$_2$]Cl around the critical endpoint. Filled and open circles represent the first-order Mott transition points where the conductance jumps and the crossover points ($P_{cross}(T),T$) defined in the text, respectively. Three arrows stand for the direction along which we evaluated the critical exponents $\delta$, $\beta$ and $\gamma$, as shown in **b–d**, respectively. **b–d**, Power-law fittings (blue lines) of the critical phenomena of conductance (see Fig. 1c and d) on logarithmic scales, where the critical values ($G_c$, $P_c$, $T_c$) ≈ (0.77 Ω$^{-1}$, 24.8 MPa, 39.7 K) are used for the analysis (Supplementary Fig. 1). The fittings give the critical exponents, ($\delta$, $\beta$, $\gamma$) ≈ (2, 1, 1).

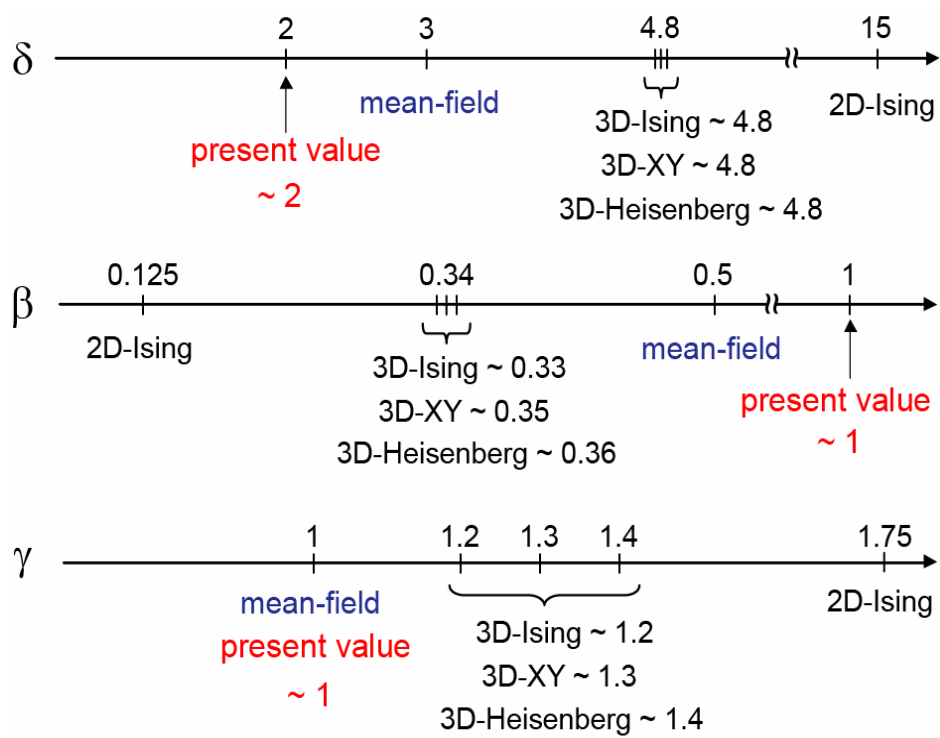

**Figure 3 | Comparison of the critical exponents ($\delta$, $\beta$, $\gamma$) of the present case with those of the known universality classes (mean-field, Ising model, XY model and Heisenberg model).** The examples of phase transitions belonging to each universality class are tabulated in Supplementary Table 1.

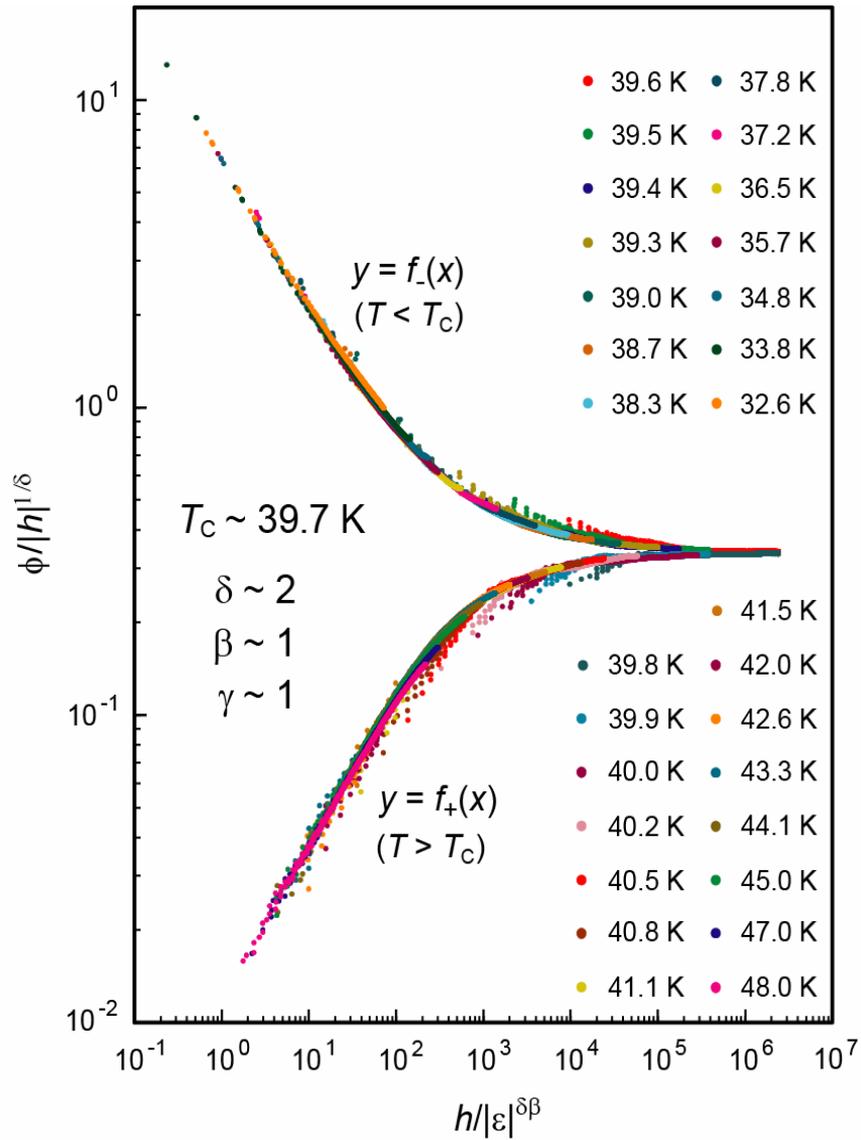

**Figure 4 | Scaling plot of the whole data set of $G_T(P)$ in the metallic region.** The plot of $\phi(h,\varepsilon)/h^{1/\delta}$ versus $h/|\varepsilon|^{\delta\beta}$ with the present values, $(\delta, \beta) \approx (2, 1)$, falls into two scaling curves corresponding to the scaling function $f_\pm(x)$ over a wide range. $[\phi, h, \varepsilon]$ in the labels of the axes corresponds to $[G_T(P) - G_T(P_{\text{cross}}(T)), P - P_{\text{cross}}(T), |T - T_c|]$ for $T > T_c$ and $[G_T(P) - G_c, P - P_1(T), |T - T_c|]$ for $T < T_c$ (see Methods).